\documentclass[amsmath,amssymb,aps,prl,twocolumn]{revtex4-1}

\usepackage{mathtools,amsmath,graphicx,units}
\usepackage{dcolumn}
\usepackage{bm}
\usepackage{xcolor}
\usepackage{physics}
\usepackage[plainpages=false,pdfpagelabels,colorlinks=true,linkcolor=red,urlcolor=blue,citecolor=blue,pdftitle={Title},pdfauthor={},pdfdisplaydoctitle=true,pdfduplex=DuplexFlipLongEdge]{hyperref}

\newcommand{\sect}[1]{\textit{#1} ---}

\newcommand{\mh}{\mathcal{H}}
\newcommand{\muu}{\mathcal{U}}
\newcommand{\mhfd}{\mathcal{H}_{\mathrm{FD}}}

\newcommand{\mbe}{\mathbf{e}}
\newcommand{\aie}[1]{a^{(#1)}_{\mbe}}

\newcommand{\bie}[1]{b^{(#1)}_{\mbe}}
\newcommand{\biep}[1]{b^{(#1)}_{\mbe^\prime}}

\newcommand{\mpn}{\langle \mathrm{PN} \rangle}
\newcommand{\mrg}{\langle r \rangle}

\newcommand{\pcsadd}{Center for Theoretical Physics of Complex Systems, Institute for Basic Science (IBS), Daejeon 34126, Republic of Korea}
\newcommand{\ustadd}{Basic Science Program, Korea University of Science and Technology (UST), Daejeon, Korea, 34113}

\begin{document}


\title{
Metal-insulator transition in infinitesimally weakly disordered flatbands
}

\author{Tilen \v{C}ade\v{z}}
    \email{tilencadez@ibs.re.kr}
    \affiliation{\pcsadd}

\author{Yeongjun Kim}
    \email{yeongjun@ust.ac.kr}
    \affiliation{\pcsadd}
    \affiliation{\ustadd}

\author{Alexei Andreanov}
    \email{aalexei@ibs.re.kr}
    \affiliation{\pcsadd}
    \affiliation{\ustadd}

\author{Sergej Flach}
    \email{sflach@ibs.re.kr}
    \affiliation{\pcsadd}
    \affiliation{\ustadd}

\date{\today}

\begin{abstract}

We study the effect of infinitesimal onsite disorder on $d$-dimensional all bands flat lattices.
The lattices are generated from diagonal Hamiltonians by a sequence of $(d+1)$ local unitary transformations parametrized by angles $\theta_i$. 
Without loss of generality, we consider the case of two flat bands separated by a finite gap \(\Delta\). 
The perturbed states originating from the flat bands are described by an effective tight binding network with finite on- and off-diagonal disorder strength which depends on
the manifold angles $\theta_i$.
The original infinitesimal onsite disorder strength \(W\) is only affecting the overall scale of the effective Hamiltonian. 
Upon variation of the manifold angles  
for \(d=1\) and \(d=2\) we find that localization persists for any choice of local unitaries, and the localization length can be maximized for specific values of $\theta_i$. 
Instead, in \(d=3\) we identify a non-perturbative metal-insulator transition upon varying the all bands flat manifold angles.

\end{abstract}

\maketitle

\sect{Introduction}
A peculiar feature of some tight-binding Hamiltonian systems is the existence of dispersionless bands known as \emph{flat bands}~\cite{derzhko2015strongly, parameswaran2013fractional, bergholtz2013topological, leykam2018artificial, leykam2018perspective}. 
Due to destructive interference and macroscopic degeneracy of such systems one can construct eigenstates which occupy only a finite number of lattice sites, called \emph{compact localized states} (CLS)~\cite{sutherland1986localization, aoki1996hofstadter}. 
The interest in flat band systems is motivated by their high sensitivity to perturbations, that can lead to various interesting physical phenomena in the presence of perturbations: 
groundstate ferromagnetism~\cite{derzhko2015strongly, tasaki1992ferromagnetism, ramirez1994strongly}, superconductivity~\cite{cao2018superconductivity},  many-body localization~\cite{kuno2020flat, danieli2020many} and unconventional Anderson localization~\cite{anderson1958absence, flach2014detangling}. 

In this Letter we consider the impact of infinitesimally weak disorder on flatbands.  
Localization in weakly disordered flat band systems was previously studied by several authors~\cite{vidal2001disorder, nishino2007flat, chalker2010anderson, leykam2013flat, flach2014detangling, leykam2017localization, ramachandran2017chiral, shukla2018disorder}. 
When some of the bands are dispersive and flatbands are gapped away from the dispersive bands, the independence of localization properties on a weak disorder strength was reported~\cite{vidal2001disorder, nishino2007flat, chalker2010anderson, leykam2013flat, leykam2017localization, shukla2018disorder} 
and effective low energy models were derived for the case of the two dimensional \(\mathcal{T}_3\) (dice) lattice~\cite{vidal2001disorder} and planar pyrochlore lattice~\cite{chalker2010anderson}. 
On the other hand, when a flat band is immersed into a dispersive band, weak disorder leads to a diverging localization length, with a variety of unconventional exponents~\cite{leykam2013flat, flach2014detangling, leykam2017localization}. 
The particular case, when a dispersive band touches a flat band, might lead to the emergence of critical states arising from the flatband states~\cite{chalker2010anderson}.

What happens when all bands turn flat -- all bands flat (ABF) systems -- where \emph{all} bands are dispersionless?
In presence of disorder there are three distinct regimes of disorder strength: 
the (infinitesimally) weak -- the disorder strength is much smaller than the gaps between the flat band energies, 
the intermediate one -- the two energy scales are comparable and the strong disorder regime. 
In the intermediate regime the flatbands hybridize due to the disorder and this can lead to delocalization and an inverse Anderson transition~\cite{goda2006inverse}, 
whereas the strong regime always leads to Anderson localization. 
For weak disorder we expect nonperturbative effects due to the interplay of the disorder and the macroscopic degeneracy of a flatband.
A metallic phase and a mobility edge around the flat band energy was numerically observed for weak disorder in a fine-tuned \(3\) dimensional \(8\) band ABF system~\cite{nishino2007flat}.

We systematically construct ABF systems in one, two and three dimensions through local unitary transformations.
We then reveal the effect of (infinitesimally) weak onsite disorder by deriving the under laying scale free effective Hamiltonian. 
We illustrate for the simplest case of two bands that infinitesimally weak disorder in one and two dimensions does not lead to delocalization.
In three dimensional systems we demonstrate a metal-insulator transition, which is driven by the variation of the parameters of the local unitary transformations.
There is in general a mobility edge in the metallic phase. 

\sect{Construction of all bands flat Hamiltonians}
A systematic way to generate ABF lattices is based on the observation that \emph{any} \(\nu\) band ABF system is represented by
a macroscopically degenerate diagonal matrix \(\mhfd\)~\cite{flach2014detangling, danieli2020nonlinear} after diagonalization. 
Since unitary transformations do not change the spectrum, \emph{any} (local or non-local) unitary transformation applied to \(\mhfd\) produces an ABF Hamiltonian with the same spectrum. 
We denote the basis where the Hamiltonian is diagonal as \emph{fully detangled} basis (FD), while the basis after the unitary transformation as \emph{fully entangled} (FE)~\cite{flach2014detangling}.
We limit our consideration to local unitary transformations (LUT) -- in the 1D case the LUT based construction is exhaustive for short-range hopping ABF Hamiltonians~\cite{danieli2020nonlinear}.
In the simplest setting LUT connect the \(\nu\) bands either in one or at most two adjacent unit cells. 
In that case the general local unitary transformations are given by SU(\(\nu\)) matrices, each of which is parametrized by \(\nu^2 - 1\) real parameters. 
In a \(d\)-dimensional system one would need at least \(d+1\) LUT in order to ensure connectivity of the final Hamiltonian. 
We denote the corresponding total unitary transformation by \(\muu\) and the transformed Hamiltonian \(\mh\) is then given as
\begin{gather}
    \mh = \muu \mhfd \muu^{\dagger}.
\end{gather}

In the examples presented below we focus on the simplest nontrivial, \(\nu = 2\) band case. 
The Hamiltonian in the FD basis is diagonal and given by
\[
    \mhfd = \Delta/2 \sum_{\mbe} \ketbra{\aie{0}}{\aie{0}} - \ketbra{\bie{0}}{\bie{0}},
\] 
where the sum runs over all the unit cells \(\mbe = n e_x\), \(\mbe = n e_x + m e_y\), \(\mbe = n e_x + m e_y + l e_z\) for \(d = 1,2,3\) and integer \(n,m,l\), respectively. 
Each LUT can then be written as
\begin{align}
    \muu_{i} = \sum_{\mbe} \left( z_i \ketbra{\aie{i}}{\aie{i-1}} + w_i \ketbra{\aie{i}}{\biep{i-1}} - \right. \notag \\
    \left. - w_i^* \ketbra{\biep{i}}{\aie{i-1}} + z_i^* \ketbra{\biep{i}}{\biep{i-1}}\right),
\end{align}
where \(i\) denotes the \(i\)-th LUT, \(\aie{i}\) and \(\bie{i}\) are the two orbitals within a unit cell \(\mbe\) in a basis after \(i\)-th LUT, 
\(\mbe^\prime = \mbe\) for \(i = 1\), \(\mbe^\prime = \mbe - e_x\) for \(i = 2\), \(\mbe^\prime = \mbe - e_y\) for \(i = 3\), 
\(\mbe^\prime = \mbe - e_z\) for \(i = 4\), \(z_i = \cos\theta_i \, e^{i \varphi_{i}}\) and \(w_i = \sin\theta_i \, e^{i \bar{\varphi}_{i}}\).

\begin{figure}
    \centering
    \includegraphics[width=0.46\textwidth]{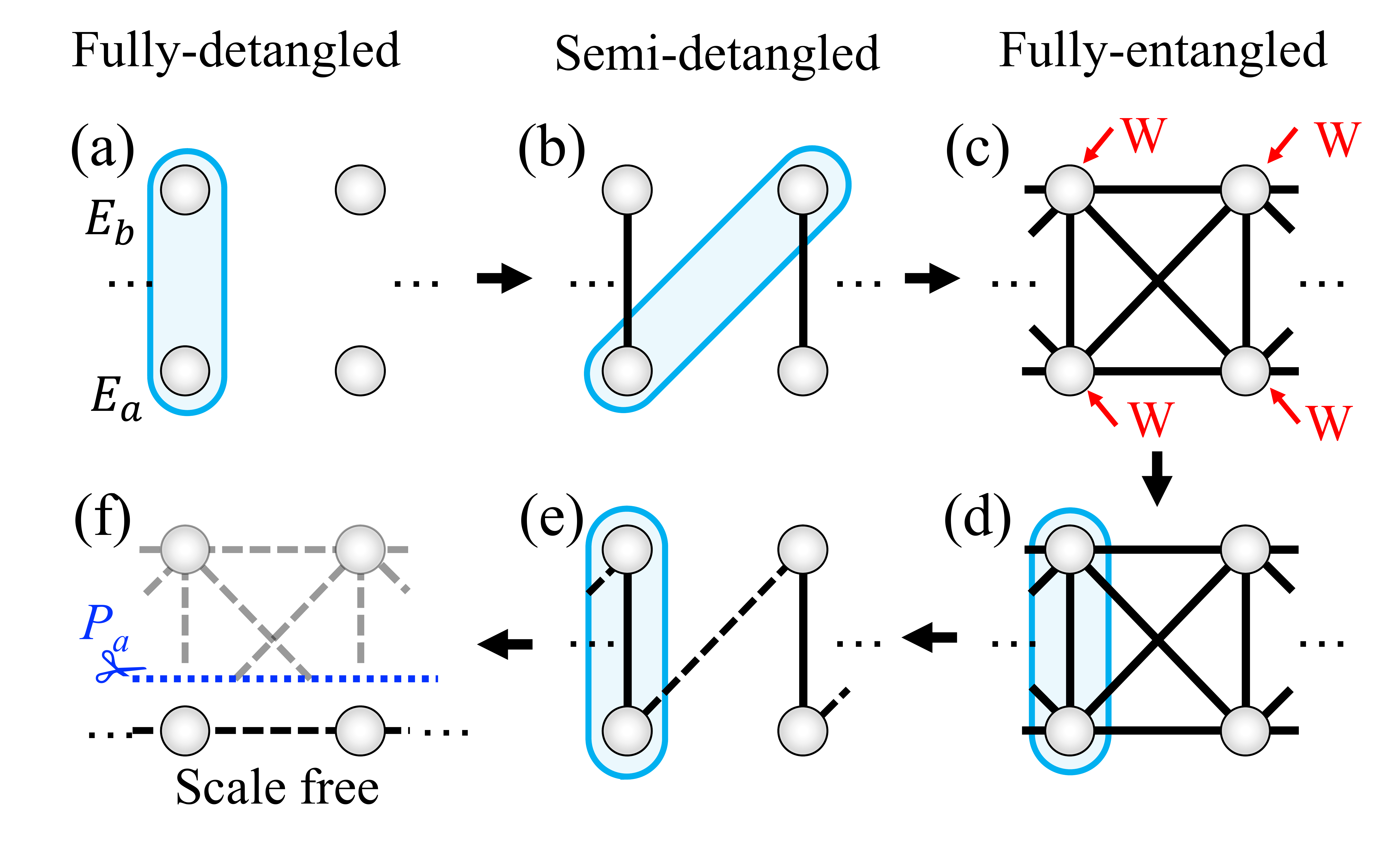} 
    \caption{
        One dimensional two band ABF system: (a) FD basis with disconnected sites. 
        First LUT, \(\muu_1\) applied within a unit cell (indicated by blue shaded area) leads to panel (b) where the second LUT, \(\muu_2\) is applied between two nearest unit cells (again indicated by the blue shaded area), leading to panel (c). 
        We apply onsite potential disorder on the sites marked by arrows and then disentangle with \(\muu_2^{\dagger}\) as indicated in panel (d). 
        This leads to panel (e) and gives rise to an effective dimerized \(1\) dimensional chain, with every second hopping element (dashed lines) being random; 
        further disentangling with \(\muu_1^{\dagger}\) brings us back to an initial FD basis, but due to disorder random hopping matrix elements of the order of disorder strength \(W\) appear within and between the nearest unit cells; 
        (f) in a weakly disordered case we project to one of the flat bands (as marked by the blue dashed line and scissors) and obtain an effective scale free single band model.
    }
    \label{Fig1}
\end{figure}

The one dimensional case construction is schematically shown in Fig.~\ref{Fig1}, panels (a-c). 
The panel (a) shows the FD basis, with circles representing disconnected sites. 
The CLS in this basis occupy a single site. 
We first apply LUT \(\muu_1\) within a unit cell as shown in panel (a) and then apply a second LUT \(\muu_2\) -- the sites affected by the LUT are indicated by blue shaded areas on the panels. 
This procedure introduces hopping matrix elements within and between the adjacent unit cells, as schematically shown in panel (c). 
In this case we then have \(\muu = \muu_2 \muu_1\).
This construction includes the well-known Creutz ladder in a magnetic field~\cite{creutz1999end} where ABF appear at half flux quantum per plaquette.
It is obtained~\cite{danieli2020nonlinear} for \(\Delta = 4 t_c\), \(\theta_1 = \theta_2 = \pi/4\), \(\varphi_1 = - \bar{\varphi}_1\) and \(\varphi_2 = \pi/2,  \bar{\varphi}_2 = 0\). 
Here \(t_c\) is the diagonal hopping element in the Creutz ladder.

In a two dimensional case the procedure is a natural extension of the one dimensional case, with the second LUT connecting the adjacent unit cells along one direction (here chosen as \(x\) direction). 
Then one applies a third LUT \(\muu_3\), which connects the adjacent unit cells along the \(y\) direction and \(\muu = \muu_3 \muu_2 \muu_1\). 
Similarly in the three dimensional case one adds the fourth LUT \(\muu_4\) to obtain \(\muu = \muu_4 \muu_3 \muu_2 \muu_1\).

\sect{Scale free model in weak disorder}
In the fully entangled basis we add a random onsite potential disorder \(\varepsilon_{\mbe, o}\) in the orbital \(o = p,f\) of the unit cell \(\mbe\) and we use the fully entangled basis labels as \(p = \aie{d+1}, f = \bie{d+1}\) -- for simplicity we consider box disorder, that is \(\varepsilon_{\mbe, o} \in \bigl[ -W/2, W/2 \bigr]\), where \(W\) is the strength of the disorder. 
Our conclusions hold for other distributions with finite support as well. 
The Hamiltonian for the onsite disorder is then \(\mh_d = W D\), where \(D\) is a diagonal matrix with \emph{all} nonzero elements drawn at random from \(\bigl[ -1/2, 1/2 \bigr]\). 
This is schematically depicted in panel (c) of Fig.~\ref{Fig1} by red arrows. 
Next we perform the inverse transformation \(\muu^{\dagger}\) into the original FD basis and write the total Hamiltonian in FD basis as \(\mh_\text{tot} = \mhfd + W \, \muu^{\dagger} D \muu\). 
We are interested in the (infinitesimally) weak disorder \(W \ll \Delta\). 
Therefore the first term \(\mhfd\) in the total Hamiltonian is much larger than the second term, while at the same time it is composed out of two flat bands. 
It follows from degenerate perturbation theory, that weak perturbations will keep the eigenvalues close to a flatband and the eigenstates will in leading order be composed of the eigenvectors from the
unperturbed Hilbert subspace which corresponds to the chosen flatband energy. 
Hence, we can project onto one of the flatbands using the projection operator \(P_a = \sum_{\mbe} \ketbra{a_{\mbe}}{a_{\mbe}}\), effectively removing the other flat band completely. 
It is worth pointing out that this projector is compact in our case~\cite{sathe2020compactly}. 
We introduce a shorthand notation \(\ket{a_{\mbe}}\equiv \ket{a^{(0)}_{\mbe}}\) for states in the FD basis. 
After the projection the \(\mhfd\) term in the total Hamiltonian becomes a trivial constant shift of energy, corresponding to the chosen flat band energy.
Thus all we are left with is the second term due to the onsite disorder. 
The projected Hamiltonian can thus be written as 
\begin{gather}
    \mh_P = W \, \mh_\text{sf},
\end{gather}
where we defined the \emph{scale free Hamiltonian} \(\mh_{\text{sf}} = P_a \muu^{\dagger} D \muu P_a\) -- the only remaining energy scale \(W\) appears just as a total prefactor.
This scale free form of the projected Hamiltonian explains the \emph{independence} of the localization properties reported in previous studies of 
flat band systems in the weak disorder regime~\cite{vidal2001disorder, nishino2007flat, chalker2010anderson, leykam2013flat, leykam2017localization, shukla2018disorder} and is one of the main results of this work.
This scale free model is the base for the analysis of the localization properties of the weakly disordered models discussed below.

\sect{Weak disorder effects in $d=1$ and $d=2$}
For the one dimensional case the inverse transformation after introducing the onsite disorder \(W\) is schematically shown in panels (d) and (e) of Fig.~\ref{Fig1}, while the projection onto the chosen flatband is depicted in panel (f). 
The scale free Hamiltonian in this case corresponds to the 1D Anderson model with correlated disorder in both onsite energies and nearest neighbor hoppings and is
\begin{gather}
    \label{eq:H_sf1d_D}
    \mh_\text{sf} = \sum_n \left( V_n \ketbra{a_n}{a_n} + t_n \ketbra{a_{n+1}}{a_n} + {\it{h.c.}} \right),
\end{gather}
where \(V_n\) and \(t_n\) are functions of LUT manifold angles and the disorder realization. 

We evaluate the impact of infinitesimal disorder on the ABF model
and study the localization properties of the scale free Hamiltonians~\eqref{eq:H_sf1d_D}. 
As a check we also make sure that the results match those obtained from the full Hamiltonian \(\mh_\text{tot}\).
We analyze the participation number PN of an eigenstate, which contains information on the localization properties of that eigenstate. 
The PN of an eigenstate \(\mu\) is defined as~\cite{wegner1980inverse, castellani1986multifractal, evers2008anderson} \(\mathrm{PN}_\mu = (\sum_{\mbe} \psi_{\mu, \mbe}^4)^{-1}\) 
where \(\psi_{\mu, \mbe}\) is the coefficient of the eigenstate in a chosen basis. 
We calculate it via the exact diagonalization (ED) in the FD real space basis (as presented in Fig.~\ref{Fig1}(f) for the 1D case). 
The localization properties are extracted from the scaling of the average PN with the system size \(\mpn \sim L^{\alpha}\), 
where \(\alpha = 0\) and \(\alpha = d\) (\(d\) being the dimension of the system) indicate localized and extended states, respectively. 

The onsite potential energies and hopping elements in~\eqref{eq:H_sf1d_D} are
\(V_n = \bigl( \varepsilon_{n,p} \cos^2\theta_2 + \varepsilon_{n,f} \sin^2\theta_2 \bigr) \cos^2\theta_1 + \bigl( \varepsilon_{n+1,f} \cos^2\theta_2 + \varepsilon_{n+1,p} \sin^2\theta_2 \bigr) \sin^2\theta_1 \) 
and 
\(t_n = 1/4 \, (\varepsilon_{n+1,f} - \varepsilon_{n+1,p}) \, \sin(2 \theta_1)\, \sin(2 \theta_2) e^{i (\varphi_1 + \bar{\varphi}_1 -\varphi_2 + \bar{\varphi}_2)}\), 
where we use the fully entangled basis labels as \(p = a^{(2)}, f = b^{(2)}\). 
We can see that the angles \(\varphi_{i}, \bar{\varphi}_{i}\) only change the overall global phase of the hoppings and can thus be set to \(0\), so that there are only two relevant LUT manifold angles left. 
Moreover, since \(\varepsilon_{n,p}\) and \(\varepsilon_{n,f}\) are uncorrelated random numbers, 
it follows from Eq.~\eqref{eq:H_sf1d_D} that disorder averaged properties are symmetric with respect to \(\theta_i \to \pi/2 - \theta_i \to \pi/2 + \theta_i\), 
thus it suffices to consider \(\theta_i \in \left[ 0, \pi/4\right]\). 
We have scanned the entire angle control parameter space and observed that the strongest enhancement of the average PN is obtained for \(\theta_1=\theta_2=\pi/4\). 
For presentation purposes we set \(\theta \equiv \theta_1 = \theta_2\).

\begin{figure}
    \centering
    \includegraphics[width=0.499\textwidth]{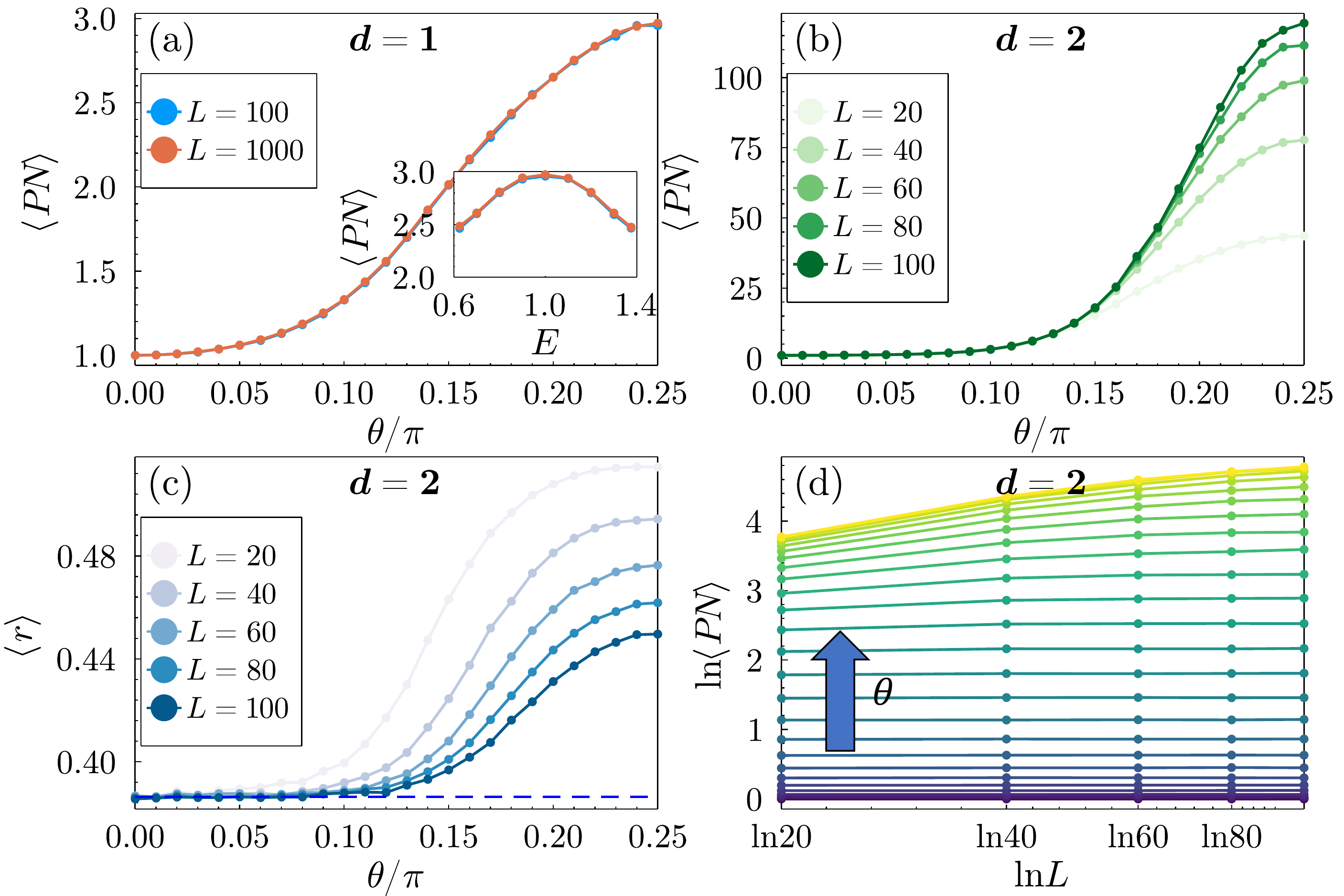} 
    \caption{
        Weakly disordered ABF in \(d=1,2\). 
        Number of unit cells used is \(L\) and \(L^2\) for \(d=1\) and \(d=2\), respectively.
        (a) Average participation number \(\mpn\) of the 1D effective model~\eqref{eq:H_sf1d_D} as a function of the manifold angle \(\theta\).
        Inset: energy dependence of \(\mpn\)  for \(\theta = \pi/4\), where the flat band energy is equal to 1. 
        All the other panels refer to the \(d=2\) case.
        (b) \(\mpn\) for the effective model as a function of \(\theta\).
        (c) Average ratio of adjacent gaps for different manifold angles \(\theta\). 
        With the increasing system size the average ROAG tends to its Poisson value, indicated by a dashed blue line.
        (d) Scaling of \(\mpn\) with the system size \(\mpn\sim L^\alpha\), \(\theta\) increases from \(0\) to \(\pi/4\) from bottom to top along the arrow direction. 
        The exponent \(\alpha \to 0\) for larger system sizes for all values of \(\theta\).
    }
    \label{Fig2}
\end{figure}

In the absence of disorder the PN of a CLS is equal to \(1\) in the FD basis. 
In contrast the average PN of the eigenstates \(\mpn\) differs significantly from this CLS result, as shown in Fig.~\ref{Fig2}(a). 
The absolute maximum is reached for \(\theta = \pi/4\), which maximizes the hopping matrix elements in Eq.~\eqref{eq:H_sf1d_D}. 
The energy dependence of the average PN shows that the largest values are obtained at the flatband energy, as shown in the inset of Fig.~\ref{Fig2}(a). 
However even in this case the average PN remains constant with the increase of the system size (\(\alpha \approx 0\)), indicating localized eigenstates~\footnote{Localization in \(d = 1\) is further confirmed by calculating the ratio of adjacent gaps, which corresponds to the Poisson distribution of energies (not shown). We have also calculated the localization length using the transfer matrix method and the results agree with the results for PN (not shown).}.

Similarly to the one dimensional case we construct the scale free model in two dimensions. 
It corresponds to the two-dimensional Anderson model with correlated onsite energies and hopping. 
The \(d = 2\) scale free model has all eigenstates being localized for all \(\theta\), although we observe much stronger finite size effects as compared to \(d=1\). 
As in \(d = 1\) we present results for reduced number of manifold angles by choosing \(\theta \equiv \theta_1 = \theta_2 = \theta_3\) and set \(\varphi_{i} = \bar{\varphi}_{i} = 0\) and observe the maximum average PN is reached for \(\theta = \pi/4\) (see Fig.~\ref{Fig2}(b)). 
The localization is confirmed by observing that the finite size scaling of the exponent \(\alpha\) of the average PN is vanishing, as can be seen in Fig.~\ref{Fig2}(d). 
We also analyze the localization properties through a spectral statistics analysis and the ratio of adjacent gaps~\cite{oganesyan2007localization, atas2013distribution} (ROAG). 
The latter is calculated from the consecutive level spacing \(s_i = E_i - E_{i-1}\), where the set of energies \(E_i\) is ordered by their ascending value. 
The ROAG is then given as \(r_i = {\mathrm{min}}(s_i, s_{i+1})/{\mathrm{max}}(s_i, s_{i+1})\) and its distributions can be compared to the random matrix theory predictions~\cite{oganesyan2007localization, atas2013distribution}. 
For ergodic systems the average ROAG \(\mrg\) is given by a value determined by the symmetry class of the system, while in the localized regime \(\mrg \sim 0.386\), corresponding to the Poisson distribution of energies. 
As shown in Fig.~\ref{Fig2}(c), we observe that the level statistics is in agreement with the Poisson distribution of energies, confirming that the eigenstates stay localized for $d=2$.

\sect{Metal-insulator transition in \(d=3\)}
As in lower dimensional cases, here we again reduce the number of LUT manifold parameters to one: \(\theta \equiv \theta_i\) for \(i=1,2,3,4\). 
The scale free Hamiltonian is a nearest neighbor and diagonal unit cell hopping cubic lattice model, which we construct and study numerically using the ED to obtain the eigenstates and eigenenergies.

\begin{figure}
    \centering
    \includegraphics[width=0.499\textwidth]{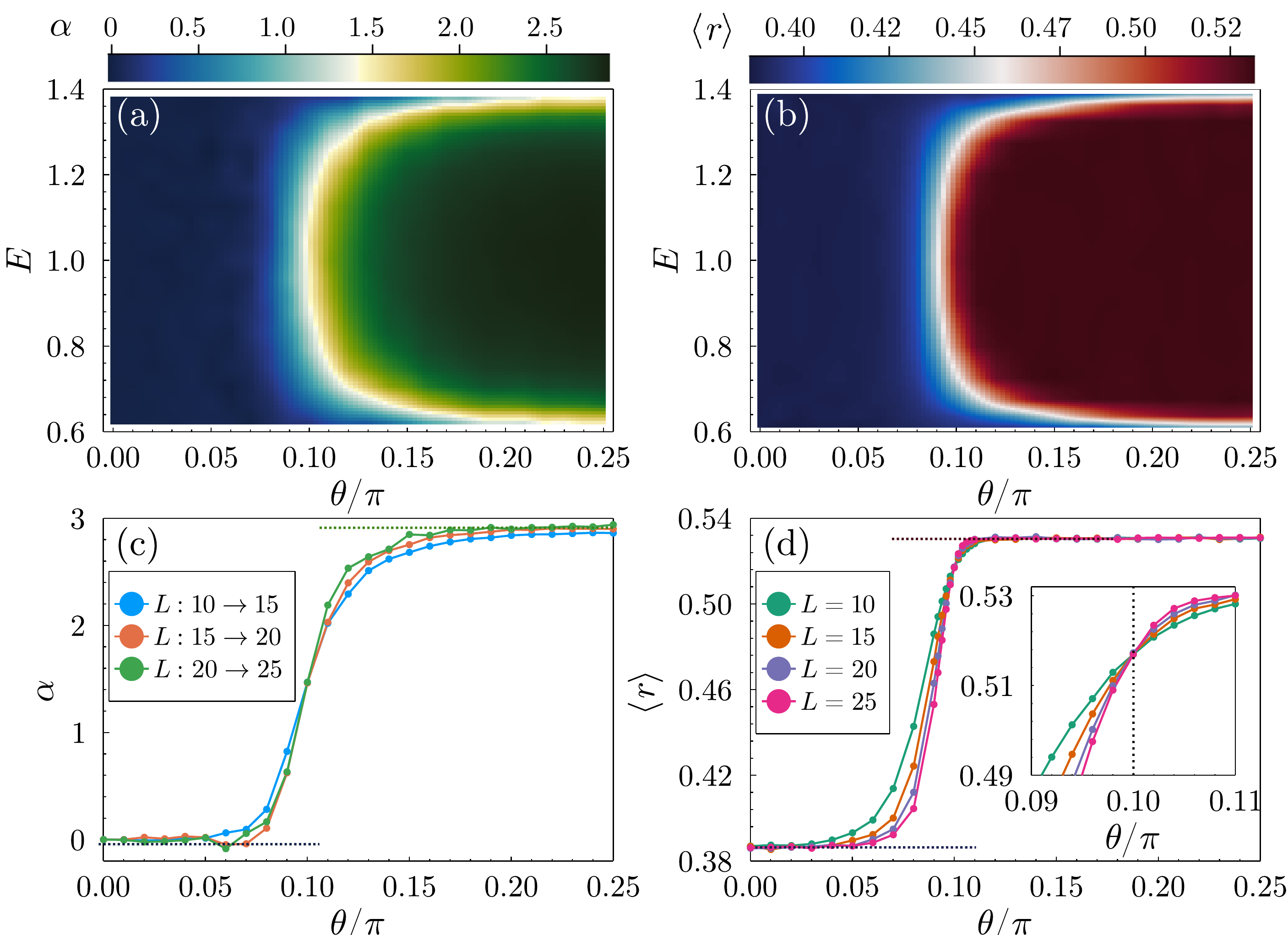} 
    \caption{
        Panels (a) and (b) show the energy resolved exponent \(\alpha\) (\(\mpn\sim L^\alpha\)) and the average ratio of adjacent gaps \(\mrg\) at weak disorder as a function of the ABF manifold angle \(\theta\). 
        Two distinct phases are clearly identified, with a white region denoting the metal-insulator transition. 
        Panels (c) and (d) show \(\alpha\) and \(\mrg\) as a function of the manifold angle \(\theta\) at the flat band energy, respectively. 
        The inset in panel (d) shows the zoom into the transition region which occurs at critical value of manifold angle $\theta_c/\pi = 0.1$.
        Number of unit cells used is \(L^3\).
    }
    \label{Fig3}
\end{figure}

We show the energy resolved average PN exponent \(\alpha\) and the average ROAG as a function of the ABF manifold angle \(\theta\) in panels (a) and (b) of Fig.~\ref{Fig3}, respectively. 
The results clearly show two distinct regions: first with \(\mpn\) not scaling with system size (\(\alpha \to 0\)) and \(\mrg \sim 0.386\), 
which correspond to localized eigenstates and Poisson distribution of energies; 
and second with \(\mpn\) increasing with the system size (\(\alpha \to d = 3\)) and \(\mrg \sim 0.531\), 
corresponding to extended eigenstates and the level statistics of the Gaussian orthogonal ensemble.
Thus we identify the former as an insulator, while the latter as a metal. 
From panels (a) and (b) of Fig.~\ref{Fig3} we also observe a mobility edge - the states at the band edge and the center of the band are localized and extended, respectively. 
Panels (c) and (d) show the exponent \(\alpha\) and the average ROAG as a function of the ABF manifold angle at the flat band energy for different system sizes, 
confirming the existence of the metal-insulator transition in weakly perturbed three dimensional ABF system, driven by varying the ABF manifold parameter.

\sect{Conclusions}
In this work we have studied all band flat systems in the presence of weak onsite potential disorder. 
Our construction of ABF systems in the continuous manifold of LUT angles enables a systematic study of effects of perturbations.  
By constructing the scale free models we explain the independence of localization properties for the weak disorder in ABF systems. 
We demonstrate that in the cases of \(d = 1\) and \(d = 2\), two band ABF systems, the eigenstates are always localized. 
In the case of \(d=3\) ABF lattices at infinitesimal disorder we observe a metal-insulator transition, driven by the change in the LUT manifold angles. 
It follows that ABF models respond qualitatively different to infinitesimally weak perturbations, allowing for phase transitions within the ABF model pool.

In addition to weak uncorrelated onsite disorder, weak randomizations of hoppings, weak random magnetic fields and fluxes, and weak hopping detuning
off the ABF model pool can be considered as additional important perturbations which can lead to novel phases and physics. In particular the
randomization of hoppings and the deviation from the ABF pool seem to be relevant issues for experimental attempts to first realize ABFs, and then to perturb them.
Additional perturbations - say slight detunings of the hoppings away from the ABF pool - will be characterized by an additional small energy scale $V$. The ratio
of that additional scale $V$ to the weak onsite disorder energy scale $W$ will be decisive. If both scales are much smaller than the FB gap $\Delta$, and if 
$W$ is still much larger than $V$, the experimental realization is expected to observe our predictions. For practical purposes it seems to suffice
to request $V/\Delta\leq 10^{-2}$ and $W/\Delta \approx 10^{-1}$.

We expect our approach to be generalizable to the case of \(\nu > 2\) all bands flat lattices, which have \(\nu\) distinct flat bands. 
Projecting to a chosen flat band should lead to an irreducible scale free model similar to the case \(\nu = 2\) discussed in this manuscript. 
More generally our results are also expected to hold for models with band structures consisting of a mix of dispersive and flat bands, in particular
for isolated \(U=1\) flatbands~\cite{flach2014detangling} with orthonormal CLS sets, that can be generated and constructed using LUT. 
For other gapped flatbands with non-orthogonal CLS sets, the scale free model can still be constructed using the projector \(P_a\) on the flat band. 
However the projector is no longer compact and shows exponentially decaying entries.  That is expected to result in exponentially decaying correlated disorder in the scale free model, which
however should not change our predictions. 
The results are expected to change qualitatively in the case of flatbands with band touchings to dispersive bands, when the CLS set ceases to be linearly independent and complete.

\sect{Acknowledgements}
T\v{C} and YK have contributed equally to this work.
T\v{C} acknowledges fruitful discussions with Carlo Danieli. 
The authors gratefully acknowledge the support by the Institute for Basic Science in Korea (Project No. IBS-R024-D1). 

\bibliography{flatband,mbl,frustration}

\end{document}